\documentclass[aps,pra,twocolumn,groupedaddress,floatfix,eqsecnum,nofootinbib,showpacs]{revtex4}
\usepackage[dvips]{graphics,color,floatflt,epsfig} 
\usepackage{amsmath}
\usepackage{amssymb}
\usepackage{graphicx} 
\usepackage{psfrag} 
\usepackage{bm} 
\input colordvi

\begin{document}

\newcommand{\Sref}[1]{Section~\ref{#1}}
\newcommand{\sref}[1]{Sec.~\ref{#1}}
\newcommand{\Cref}[1]{Chap.~\ref{#1}}
\newcommand{\tql}{\textquotedblleft} 
\newcommand{\tqr}{\textquotedblright~} 
\newcommand{\tqrc}{\textquotedblright} 
\newcommand{\Refe}[1]{Equation~(\ref{#1})}
\newcommand{\Refes}[1]{Equations~(\ref{#1})}
\newcommand{\fref}[1]{Fig.~\ref{#1}}
\newcommand{\frefs}[1]{Figs.~\ref{#1}}
\newcommand{\Fref}[1]{Figure~\ref{#1}}
\newcommand{\Frefs}[1]{Figures~\ref{#1}}
\newcommand{\reff}[1]{(\ref{#1})}
\newcommand{\refe}[1]{Eq.~(\ref{#1})}
\newcommand{\refes}[1]{Eqs.~(\ref{#1})}
\newcommand{\refi}[1]{Ineq.~(\ref{#1})}
\newcommand{\refis}[1]{Ineqs.~(\ref{#1})}
\newcommand{\framem}[1]{\overline{\overline{\underline{\underline{#1}}}}}
\newcommand{\PRA }{{ Phys. Rev.} A }
\newcommand{\PRB }{{ Phys. Rev.} B} 
\newcommand{\PRE }{{ Phys. Rev.} E}
\newcommand{\PR}{{ Phys. Rev.}} 
\newcommand{\APL }{{ Appl. Phys. Lett.} }
\newcommand{\PRL}{Phys.\ Rev.\ Lett. }
\newcommand{\OCOM }{{ Opt. Commun.} } 
\newcommand{\JOSA }{{ J. Opt. Soc. Am.} A}
\newcommand{\JOSB }{{ J. Opt. Soc. Am.} A}
\newcommand{\JMO }{{J. Mod. Opt.}}
\newcommand{\RMP}{Rev. \ Mod. \ Phys. }
\newcommand{\etal} {{\em et al.}}

\title{Optical cavities and waveguides in hyperuniform disordered photonic solids}
\author{Marian Florescu$^{1}$} \email[Electronic Address: ]{ m.florescu@surrey.ac.uk}
 \author{Paul J. Steinhardt$^{2,3}$} \author{Salvatore Torquato$^{2,3,4}$}

\affiliation{$^1$ Advanced Technology Institute and Department of Physics, University of
  Surrey, Surrey, United Kingdom} \affiliation{$^2$ Department of Physics, Princeton
  University, Princeton, New Jersey, 08544, USA} \affiliation{$^3$Princeton Center for
  Theoretical Science, Princeton University, Princeton, New Jersey 08544, USA}
\affiliation{$^4$Department of Chemistry, Princeton University, Princeton, New Jersey
  08544, USA}

\date{\today}

\begin{abstract}
  Using finite difference time domain and band structure computer simulations, we show
  that it is possible to construct optical cavities and waveguide architectures in
  hyperuniform disordered photonic solids that are unattainable in photonic crystals.  The
  cavity modes can be classified according to the symmetry (monopole, dipole, quadrupole,
  etc.) of the confined electromagnetic wave pattern.  Owing to the isotropy of the band
  gaps characteristic of hyperuniform disordered solids, high-quality waveguides with
  freeform geometries (e.g., arbitrary bending angles) can be constructed that have no
  analogue in periodic or quasiperiodic solids.  These capabilities have implications for
  many photonic applications.
\end{abstract}

\pacs{41.20.Jb, 42.70.Qs, 78.66.Vs, 85.60.Jb}
\maketitle

Recently, we introduced \tql hyperuniform stealthy\tqr disordered photonic
solids with large isotropic band gaps comparable in width to the
anisotropic band gaps found in photonic crystals and capable of blocking
light of all polarizations Ref.~\cite{pnas_flo}. These solids challenge the
conventional wisdom that band gaps require Bragg scattering and,
hence, periodic or quasiperiodic order.  The hyperuniform solids
described in [1] demonstrate that Mie scattering is sufficient to
generate band gaps provided the disorder is constrained to be
hyperuniform (see definition below).  We have explained how to design
the dielectric materials in Ref. \cite{pnas_flo} and explored their
band-gap and transport properties in Ref. \cite{sub_flo_1}.

\begin{figure}[h]
  {\centerline{ \includegraphics*[width=0.9\linewidth]{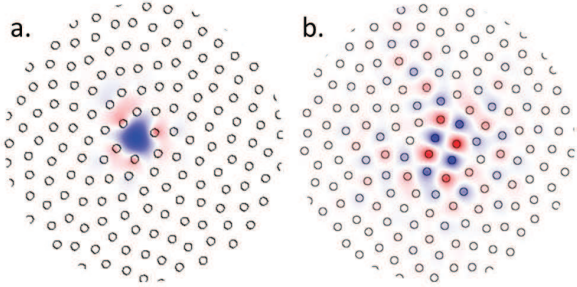}}}
\caption{ (Color online) Electric field distribution for  a) a confined cavity mode in a
  hyperuniform disordered photonic structure introduced into the band gap by removing a
  dielectric cylinder from the original structure; to be compared with b) a localized mode
  in the defect-free photonic structure.  Note that the cavity mode has a shorter
  localization length. } \label{fig_cav_s}
\end{figure}

In this Letter, we explore the cavity and waveguide architectures possible in hyperuniform
disordered solids.  We find a wide range of confined cavity modes characterized by
different approximate symmetries (monopole, dipole, quadrupole, etc.) and high-quality
freeform waveguides that are possible because of the intrinsic isotropy of the solids and
their band gaps.

Central to the class of materials considered in this paper is the
concept of hyperuniformity, which was first introduced as an order
metric for ranking point patterns according to their local density
fluctuations at large length scales \cite{torquato_2003}. A point
pattern in real space is hyperuniform if the number variance
$\sigma(R)^2$ within a spherical sampling window of radius $R$ (in $d$
dimensions), grows more slowly than the window volume for large $R$,
i.e., more slowly than $R^d$.  Crystalline and quasicrystalline point
patterns trivially satisfy this property, but it is also possible to
have isotropic, disordered hyperuniform point patterns.  In Fourier
space, hyperuniformity means the structure factor $S({\bf k})$
approaches zero as $|{\bf k}| \rightarrow 0$.  The hyperuniform
patterns that we consider are restricted to the subclass in which the
number variance grows like the window surface area for large $R$,
e.g., $\sigma^2(R)=A R$ in two-dimensions, or $\sigma^2(R)=A R^2$ in
three dimensions, up to small oscillations
\cite{torquato_2003,sal_Acoeff}. 

We further constrain the disorder to produce hyperuniform {\it
  stealthy} point patterns for which the structure factor $S({\bf k})$
is isotropic and precisely equal to zero for a finite range of
wavenumbers $0 \le k \le k_C$ for some positive critical wavevector,
$k_C$ \cite{hyper_stealthy}.  Hyperuniform photonic materials are then
constructed by decorating a hyperuniform stealthy point pattern with
dielectric materials according to the protocol described in
Ref.~\cite{pnas_flo}.  As a result of the constrained disorder, the HD
photonic materials display an unusual combination of physical
characteristics.  Some are associated with typical disordered
structures, such as statistical isotropy and multiple scattering
resulting in localized states.  Others, such as the existence of large
and robust band gaps, result from a combination of hyperuniformity,
uniform local topology (e.g., in 2d, a network of vertices in which
all connections are trivalent), and short-range geometric order
(derived from the stealthiness) \cite{pnas_flo}.  Following our
protocol, dielectric heterostructures with large, complete (both
polarizations) band gaps have been designed.  Recently, these designs
have been fabricated on the microwave scale and successfully tested
\cite{Man2010}.

The structures analyzed in this paper are generated by decorating
hyperuniform point patterns with cylindrical rods with dielectric
constant $\epsilon=11.56$ and radius $r/a=0.189$; these values are
chosen to optimize the size of the photonic band gap.  Here, we have
introduced a length scale $a = L/\sqrt{N}$ , such that the
hyperuniform pattern has density of $1/a^2$. The hyperuniform point
patterns are generated using the collective coordinate method in
\cite{torquato_2003} with stealthy order parameter $\chi=0.5$.  It is
notable that the photonic band gaps (PBGs) for these disordered
structures are equivalent to the fundamental band gap in periodic
systems, i.e., the spectral location of the gap is determined by the
resonant frequencies of the scattering centers and always occurs
between band $N$ and $N + 1$, with $N$ precisely the number of points
per unit cell. A typical PBG size for structures with $\chi=0.5$ is
$\Delta \omega / \omega_C=37\%$, where $\omega_C$ is the central
frequency of the gap.  Here, for simplicity, we consider PBGs for
transverse magnetic (TM) polarized radiation.

We use the finite-difference time-domain (FDTD) method \cite{fdtd_1}
to calculate the propagation of light inside the HD photonic
structures \cite{sub_flo_1}. We employ a computational domain with
periodic boundary conditions in the transverse direction and perfectly
matched layer (PML) condition in the normal direction.  The spatial
resolution in our numerical experiments is at least $n=64$ mesh points
per $a$, and the temporal resolution is $0.5/n \times a/c$, where $c$
is the light speed in vacuum. For transmission calculations, a
broadband source is placed at one end of the computational domain and
the transmission signal is recorded at the other end with a
line-detector \cite{fdtd_2}. The Fourier components of the field are
then evaluated and the spectra are averaged and normalized to the
transmission profile in the absence of the structure.  For quality
factor calculations \cite{quality_factor}, the modes are excited with
a broadband pulse from a current placed directly inside the cavity and
the simulation domain is surrounded by PML all around.  After the
source is turned off, the fields are analyzed, and frequencies and
decay rates of the confined modes are evaluated \cite{fdtd_2}. To
calculate photonic band structures, we employ a supercell
approximation and use the conventional plane-wave expansion method
\cite{bs_1,bs_2}. In all the simulations performed in
this work, the computational domain size is $\sqrt{N} a \times
\sqrt{N} a$, with $N=500$.

\begin{figure}[ht]
{\centerline{ \includegraphics*[width=0.9\linewidth]{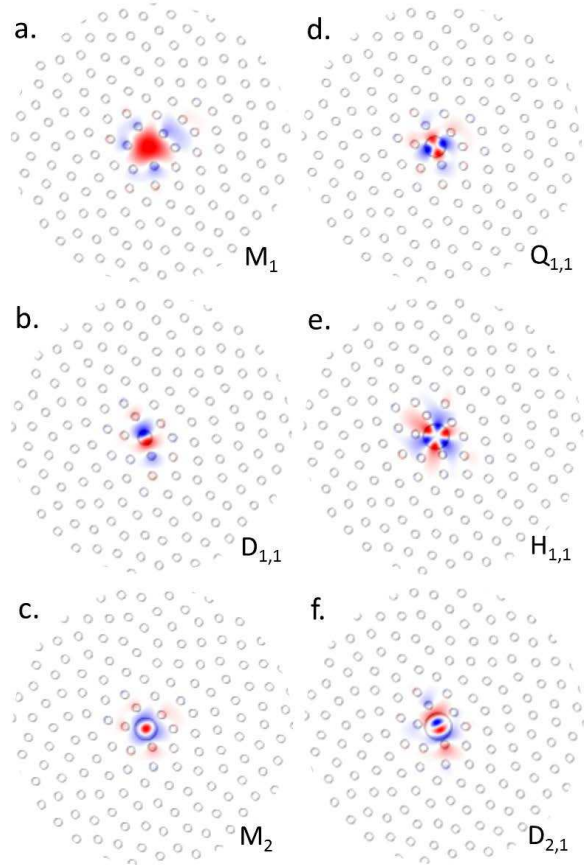}}}
\caption{ (Color online) Electric field distribution for various cavity modes (defined in
  the text) obtained for different radii of the perturbed cylinder. From a) to f), the
  dimensionless defect radius $r_d/r_0$ takes the values: $0,1.8,2.7,2.4,3.0,3.4$,
  respectively. } \label{fig_cav_fields}
\end{figure}

\begin{figure}[ht]
{\centerline{ \includegraphics*[width=0.9\linewidth]{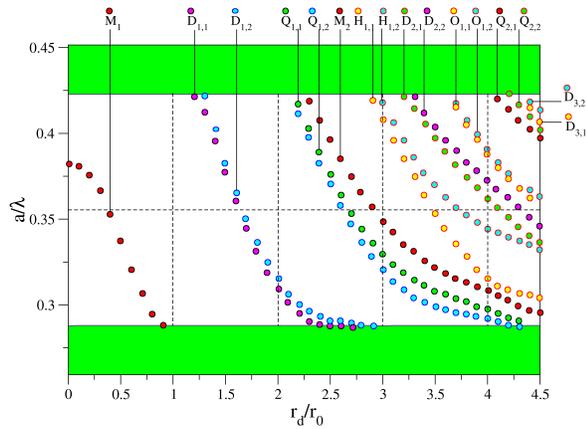}}}
\caption{ (Color online) Photonic band structure calculations showing
  the evolution of localized modes associated with a defect cylinder
  as a function of the dimensionless defect radius $r_d/r_0$ (where
  $r_d/r_0=1$ corresponds to the unperturbed HD photonic
  structure). Different localized modes are labeled based on their
  approximate symmetry an order (e.g. $D_{1,2}$ indicates the second
  mode of first order with a dipole-like symmetry).  The (green)
  shaded regions represent the continuous of modes that bound the PBG
  frequency range. For $r_d/r_0 < 1$ (the radius of the cylinder is
  decreased) only a single mode with a \tql monopole\tqrc is pushed up
  into the gap.  For $r_d/r_0 > 1$ (the radius of the cylinder is
  increased), higher-order modes are descending into the PBG. The
  electric field patterns for some of these modes are shown in
  \fref{fig_cav_fields}.} \label{fig_cav_modes}
\end{figure}

In an otherwise unperturbed HD structure, it is possible to create a
localized state of the electromagnetic field by reducing or enhancing
the dielectric constant at a certain point in the sample. In the
two-dimensional structures considered here, this can be realized by
removing one of the cylinders. Due to the presence of the point-like
defect, a localized cavity mode is created within the photonic band
gap at a certain frequency. \Fref{fig_cav_s}a shows a cavity mode
obtained by removing one of the dielectric cylinders from a HD
structure. We note that the electric field distribution is highly
localized around the defect, extending only up to distances involving
1-2 rows of cylinders beyond the position of the missing cylinder. The
quality factor of the two-dimensional confined mode is higher than
$10^{8}$. It is expected the for three-dimensional slab structures
obtained by slab-configurations with a fine thickness, quality factors
of at least $10^3$ can be maintained, similar to the case of cavities
in quasiperiodic photonic structures \cite{quality_factor}.  The
nature of the localization mechanism around this type of defect in HD
materials is rather different from the Anderson-like localization
mechanism naturally present in this as well as conventional disordered
structures.  \Fref{fig_cav_s} shows a localized photonic mode in the
unperturbed HD structure has a localization length that is 5-6 times
larger than that in the cavity mode.

\begin{figure}[ht]
{\centerline{ \includegraphics*[width=0.9\linewidth]{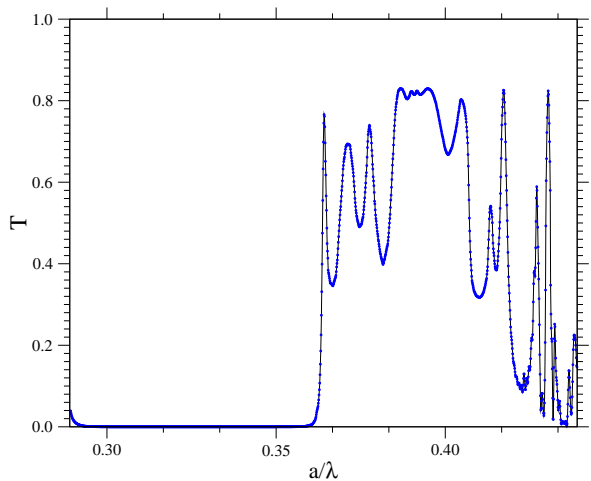}}}
{\caption{ (Color online) Transmission spectrum of a guided mode in a
hyperuniform disordered photonic structure created by removing
dielectric cylinders along a sinusoidally-shaped path.}}
\label{fig_wg_s}
\end{figure}

\begin{figure}[ht]
{\centerline{ \includegraphics*[width=0.9\linewidth]{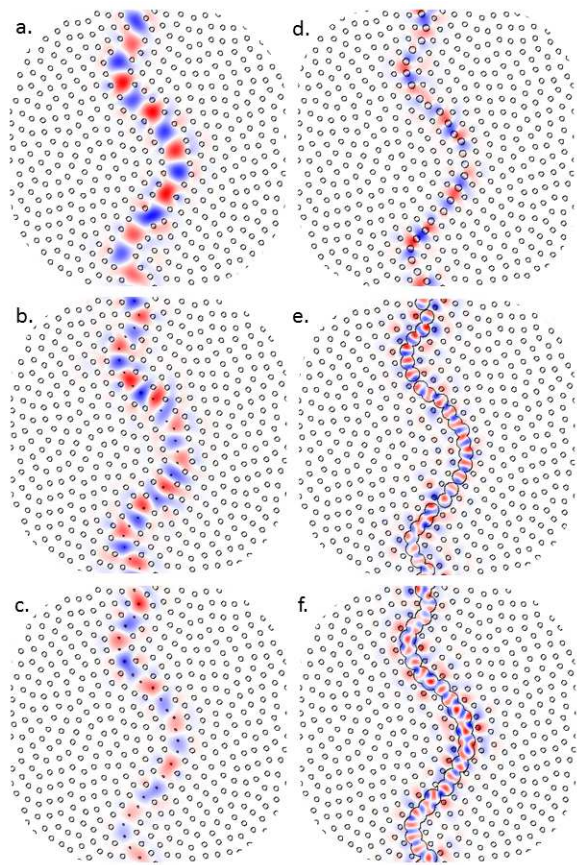}}}
\caption{ (Color online) Electric field distribution for various guided modes
  obtained for different radii of the perturbed cylinders. From a) to
  f), the dimensionless defect radius $r_d/r_0$ takes the values:
  $0,0.3,0.4,1.4,3.0,3.4$, respectively.} \label{fig_wg}
\end{figure}

We next study the evolution of the localized modes associated with a
perturbed cylinder as its radius varies.  When the radius of the
cylinder is reduced, a single mode from the continuum of modes below
the lower photonic band edge is pulled inside the PBG and becomes
localized.  If the radius of the cylinder is increased, a number of
modes (the precise number is determined by the relative size of the
defect cylinder) from the continuum of modes above the upper photonic
band edge are pulled inside the PBG. \Fref{fig_cav_fields} shows the
electric field mode distribution for a few selected localized
modes. Note the nearly perfect monopole (M), dipolar (D), quadrupolar
Q), and hexapolar (H) symmetries associated with certain
modes. Different localized modes are indexed based on their
approximate symmetry (M,D,H,...), where the first index refers to the
order of the mode and the second index refers to the number of modes
of a given order (e.g., $D_{1,2}$ is the second mode of first-order
with a dipole-like symmetry).

In \fref{fig_cav_modes} we show the evolution of the localized modes
associated with a defect cylinder as a function of the dimensionless
defect radius.  Let us define the dimensionless defect radius to be
$r_D/r_0$. For a defect radius $r_d/r_0=0.47$ (where $r_0$ is the
radius of the unperturbed cylinders), the defect mode reaches the
mid-point of the PBG and is maximally protected from interactions with
the propagating modes from the continua below and above the photonic
band gap. When the radius of the defect cylinder is increased, it
becomes possible to accommodate more localized modes in the defect
region, distinguished either by their approximate symmetry or/and
frequency.  For $r_d/r_0=4$, a total of 12 localized modes can coexist
within the same defect. However, it should be noted that at at these
large radii, the defect cylinders start to overlap with the
surrounding cylinders and the confinement decreases.

We now consider waveguide architectures. In photonic crystals,
removing a row of rods generates a channel through which light with
frequencies within the band gap can propagate, a so-called crystal
waveguide.  Light cannot propagate elsewhere in the structure outside
the channel because there are no allowed states.  The waveguides must
be composed of segments whose orientation is confined to the
high-symmetry directions of the crystal. As a result, the waveguide
bends of 60$^{\circ}$ or 90$^{\circ}$ can be easily achieved, but
bends at an arbitrary angle lead to significant radiation loss due to
excessively strong scattering at the bend junction and require
additional engineering to function properly.

The existence of large and robust photonic band gaps in HD structures
suggests that waveguiding should be possible in these non-crystalline
photonic solids. An important difference is that the distribution of
dielectric material around the bend junction is statistically
isotropic. If the defect mode created by the removal of material falls
within the PBG, the bend can then be oriented at an arbitrary angle.
In \fref{fig_wg} a) an example of a guided mode obtained by removing
dielectric cylinders along a sinusoidally-shaped path through the HD
structure. Remarkably, the light propagating through this
unusually-shaped waveguide channel is tightly confined in the
transverse direction, penetrating only in the next few rows of
dielectric cylinders. Our calculations show that the transmission
reaches a maximum of about $83\%$ see \fref{fig_wg_s}. It is well
known that in a photonic crystal, conservation of momentum due to the
translation invariance along a linear waveguide prevents back
scattering of the propagating mode. Such a mechanism is absent in any
waveguide that presents deviations from linearity be it in a periodic
or disordered structure, but it can be alleviated by optimizing the
cylinder size along the waveguide channel.

The hyperuniform disordered structures analyzed here yield large
photonic band gaps of around 40\% of the central frequency, which in
turn suggests that higher-order guided modes can be excited in an
appropriately designed waveguide channel. In \fref{fig_wg}, we also show
higher-order guided modes that are obtained by varying the radius of
the defect cylinders along the channel path.

In summary, we have introduced novel architectures for the design of
optical cavities and waveguides in hyperuniform disordered
materials. We have demonstrated that point-like defects can support
localized modes with a variety of symmetries and multiple frequencies.
By exploiting the isotropy of the PBG unique to
hyperuniform disordered structures, we have also shown that it is possible to 
design waveguides of essentially arbitrary shape, along which the
light can be guided through the excitation of localized resonances
similar to the ones that we found in the point-like defects.  The
ability to localize modes of different symmetry and frequency in the
same physical cavity and to guide light through modes with different
localization properties can have a great impact on
all-optical switching and single-atoms laser systems
\cite{eu_switch,lucia_atom}. The new cavity and waveguide
architectures are promising candidates for achieving highly flexible
and robust platforms for integrated optical micro-circuitry.

While the present study deals only with TM polarized radiation,
qualitatively similar results can be obtained for transverse electric
(TE) polarized radiation for structures designed using the constrained
optimization protocol developed in Ref. \cite{pnas_flo}. This will be
the subject of a future investigation.

\begin{acknowledgments}
This work was supported by National Science Foundation under Grants No. DMR-0606415 and
ECCS-1041083.
\end{acknowledgments}

\end{document}